\documentclass[twocolumn,superscriptaddress,prb,showpacs,floatfix]{revtex4}
\newif\ifGALLEYversion\GALLEYversionfalse

\usepackage{ifthen}
\usepackage{graphicx}
\usepackage{amsmath}
\usepackage{braket}
\usepackage{bm}

\ifthenelse{\boolean{GALLEYversion}}{
   \marginparwidth 2.7in
   \marginparsep 0.5in
   \def\ttm#1{\marginpar{\small TT: #1}}
   \def\glm#1{\marginpar{\small GL: #1}}
   \def\nhm#1{\marginpar{\small NM: #1}}
   \def\dcm#1{\marginpar{\small DC: #1}}
}{
   \def\ttm#1{\relax}
   \def\glm#1{\relax}
   \def\nhm#1{\relax}
   \def\dcm#1{\relax}
}

\newcommand{\beq}{\begin{equation}}
\newcommand{\eeq}{\end{equation}}
\newcommand{\bea}{\begin{eqnarray}}
\newcommand{\eea}{\end{eqnarray}}
\newcommand{\nn}{\nonumber\\}
\newcommand{\rvec}{{\bf r}}
\newcommand{\Rvec}{{\bf R}}
\newcommand{\Avec}{{\bf A}}
\newcommand{\Bvec}{{\bf B}}
\newcommand{\pvec}{{\bf p}}

\newcommand{\vvec}{{\bf v}}
\newcommand{\kvec}{{\bf k}}
\newcommand{\Lvec}{{\bf L}}
\newcommand{\mvec}{{\bf m}}
\newcommand{\Mvec}{{\bf M}}

\newcommand{\evec}{{\bf e}}

\newcommand{\tauvec}{\ensuremath{\bm{\tau}}}
\newcommand{\Hcal}{\ensuremath{\mathcal{H}}}

\newcommand{\VRNL}{V_{\bf R}^\mathrm{NL}}
\newcommand{\KRNL}{K_{\bf R}^\mathrm{NL}}
\newcommand{\ERNL}{{\bf E}_{\bf R}^\mathrm{NL}}

\newcommand{\ptilde}{\ensuremath{\widetilde{p}}}
\newcommand{\psitilde}{\ensuremath{\widetilde{\psi}}}
\newcommand{\phitilde}{\ensuremath{\widetilde{\phi}}}
\renewcommand{\AE}{\ensuremath{\mathrm{AE}}}
\newcommand{\PS}{\ensuremath{\mathrm{PS}}}
\newcommand{\GIPAW}{\ensuremath{\mathrm{GIPAW}}}
\newcommand{\exponent}{(i/2c)\rvec\cdot\Rvec\times\Bvec}

\begin{document}
\title{An ab-initio converse NMR approach for pseudopotentials}
\author{D. Ceresoli}
\affiliation{Department of Materials Science and Engineering,
MIT, Cambridge, Massachusetts 02139, USA.}
\author{N. Marzari}
\affiliation{Department of Materials Science and Engineering,
MIT, Cambridge, Massachusetts 02139, USA.}
\affiliation{University of Oxford, Department of Materials,
Parks Road, Oxford OX1 3PH, UK.}
\author{M. G. Lopez}
\affiliation{Department of Physics, Wake Forest University,
Winston-Salem, North Carolina 27109, USA.}
\author{T. Thonhauser}
\affiliation{Department of Physics, Wake Forest University,
Winston-Salem, North Carolina 27109, USA.}
\date{\today}

\begin{abstract}
We extend the recently developed converse NMR approach [T. Thonhauser,
D. Ceresoli, A. Mostofi, N. Marzari, R. Resta, and D.  Vanderbilt,
J. Chem. Phys. \textbf{131}, 101101 (2009)] such that it can be used
in conjunction with norm-conserving, non-local pseudopotentials. This
extension permits the efficient ab-initio calculation of NMR chemical
shifts for elements other than hydrogen within the convenience of a
plane-wave pseudopotential approach. We have tested our approach on
several finite and periodic systems, finding very good agreement with
established methods and experimental results.
\end{abstract}
\maketitle

\section{Introduction}
The experimental technique of nuclear magnetic resonance (NMR) is a
powerful tool to determine the structure of molecules, liquids, and
periodic systems. It is thus not surprising that, since its discovery in
1938, NMR has evolved into one of the most widely used methods in
structural chemistry.\cite{Rabi,NMR_encyclopedia} Unfortunately, one
caveat of this successful method is that there is no basic, generally
valid ``recipe'' that allows a unique determination of the structure
given a measured spectrum. As a result, for more complex systems the
mapping between structure and measured spectrum can be ambiguous.

It had been realized early on that ab-initio calculations could resolve
some of these ambiguities and thus greatly aid in determining structures
from experimental NMR spectra. For finite systems such as simple
molecules, appropriate methods were first developed in the
quantum-chemistry community.\cite{Kutzelnigg_90} While highly accurate,
these methods by construction were unable to calculate NMR shifts of
periodic systems, which is important for the increasingly popular
solid-state NMR spectroscopy. The underlying physical limitation is due
to the fact that the description of any constant external magnetic field
requires a non-periodic vector potential. Possible approaches to combine
such a non-periodic vector potential with the periodic potential of
crystals were found only within the last decade.\cite{Mauri_96,
Sebastiani_01, Pickard_Mauri_01,Pickard_Mauri_03,Yates_07} All of these
approaches have in common that they treat the external magnetic field in
terms of the linear-response it causes to the system under
consideration. Although these approaches are accurate and successful,
the required linear-response framework makes them fairly complex and
difficult to implement.

Recently, a fundamentally different approach for the calculation of
ab-initio NMR shifts has been developed by some of
us.\cite{Thonhauser_09} In our converse approach we circumvent the need
for a linear-response framework in that we relate the shifts to the
macroscopic magnetization induced by magnetic point dipoles placed at
the nuclear sites of interest. The converse approach has the advantage
of being conceptually much simpler than other standard approaches and it also
allows us to calculate the NMR shifts of systems with several hundred
atoms. Our converse approach has already successfully been applied to
simple molecules, crystals, liquids, and extended polycyclic aromatic
hydrocarbons.\cite{Thonhauser_08,Thonhauser_09}

While the converse method can be directly implemented in an
all-electron first-principles computer code, an implementation into a
pseudopotential code is more complicated due to nonlocal projectors
usually used in the Kleinman-Bylander separable form.\cite{Kleinman_82}
In this paper we present a mathematical extension to the converse
formalism such that it can be used in conjunction with norm-conserving,
non-local pseudopotential. This extension permits the efficient
ab-initio calculation of NMR chemical shifts for elements other than
hydrogen within the convenience of a plane-wave pseudopotential
approach.

The paper is organized in the following way: In Sec.~\ref{sec:theory}
we first review the converse-NMR method and its relation to the
orbital magnetization at an all-electron level. Then, we apply the
gauge including projector augmented wave (GIPAW)
transformation to derive an expression for the orbital
magnetization in the context of norm-conserving pseudopotentials.
We discuss aspects of the implementation of the converse method in
Sec.~\ref{sec:implementation}. In this section we also show results
of several convergence tests that we have performed.  To validate
our approach, we apply our converse approach to molecules and
solids and the results are collected in Sec.~\ref{sec:results}. In
Sec.~\ref{sec:discussion} we discuss the main advantages of the converse
method. Finally, we summarize and conclude in Sec.~\ref{sec:summary}. The
GIPAW transformation and several details of the mathematical
formalism---which would be distracting in the main text---are presented in
Appendices~\ref{app:gipaw} and \ref{app:periodic}.

\section{Theory}\label{sec:theory}
The converse method for calculating the NMR chemical shielding has been
introduced in Ref.~[\onlinecite{Thonhauser_09}] and can be summarized as
follows:
\beq\label{eq:converse}
  \sigma_{s,\alpha\beta} = \delta_{\alpha\beta}
  -\Omega\frac{\partial M_\beta}{\partial m_{s,\alpha}}
\eeq
Thus, in the converse method the chemical shielding tensor
$\sigma_{s,\alpha\beta}$ is obtained from the derivative of the orbital
magnetization $\Mvec$ with respect to a magnetic point dipole $\mvec_s$,
placed at the site of atom $s$. $\delta_{\alpha\beta}$ is the Kronecker
delta and $\Omega$ is the volume of the simulation cell. In other words,
instead of applying a constant magnetic field to an infinite periodic
system and calculating the induced field at all equivalent $s$ nuclei,
we apply an infinite array of magnetic dipoles to all equivalent sites
$s$, and calculate the change in magnetization.  Since the perturbation
is now periodic, the original problem of the non-periodic vector potential
has been circumvented.

In practice, the derivative in Eq.~(\ref{eq:converse}) is calculated as
a finite difference of the orbital magnetization in presence of a small
magnetic point dipole $\mvec_s$. Since $\Mvec$ vanishes for $\mvec_s =
0$ and is an odd function of $\mvec_s$ because of time-reversal symmetry
(for a non-magnetic system in absence of spin-orbit interaction),
it is sufficient to perform three calculations of $\Mvec(m_s\,\evec)$,
where $\evec$ are cartesian unit vectors.

Within density-functional theory (DFT) the all-electron Hamiltonian is,
in atomic units:
\beq\label{eq:Hae_conv}
  \Hcal = \frac{1}{2}\left(\pvec +\frac{1}{c}\Avec_s(\rvec)\right)^2 +
    V_\mathrm{KS}(\rvec)
\eeq
where
\beq\label{eq:A_s}
  \Avec_s(\rvec) = \frac{\mvec_s \times(\rvec - \rvec_s)}{|\rvec-\rvec_s|^3}
\eeq
is the vector potential corresponding to a magnetic dipole $\mvec_s$
centered at the atom $s$ coordinate $\rvec_s$.~\cite{Jackson}
We neglect any explicit dependence of the exchange-correlation functional
on the current density. In practice, spin-current density-functional
theory calculations have shown to produce negligible corrections
to the orbital magnetization.~\cite{sharma07}

In finite systems (i.e. molecules), the orbital magnetization can be
easily evaluated via the velocity operator $\vvec = -i[\rvec,\Hcal]$:
\beq\label{eq:M_velocity}
  \Mvec = \frac{1}{2c}\sum_n\braket{\psi_n|\rvec\times\vvec|\psi_n}
\eeq
where $\ket{\psi_n}$ are molecular orbitals, spanning the occupied
manifold.  For periodic systems, the situation is more complicated
due to itinerant surface currents and to the incompatibility of the
position operator $\rvec$ with periodic boundary conditions.  It has
been recently shown~\cite{Resta_05,Thonhauser_06,Ceresoli_06,Niu} that
the orbital magnetization in a periodic system is given by:
\begin{multline}\label{eq:M_periodic}
  \Mvec = -\frac{1}{2c}\mathrm{Im} \sum_{n\kvec} f(\epsilon_{n\kvec})\bra{
  \partial_\kvec u_{n\kvec}}\\
  \times\left(\Hcal_\kvec + \epsilon_{n\kvec} -
  2\epsilon_\mathrm{F}\right)\ket{\partial_\kvec u_{n\kvec}}
\end{multline}
where $u_{n\kvec}$ are the Bloch wavefunctions, \mbox{$\Hcal_\kvec
= e^{-i\kvec\rvec}\; \Hcal\;e^{i\kvec\rvec}$}, $\epsilon_{n\kvec}$
are its eigenvalues and $\epsilon_\mathrm{F}$ is the Fermi level.
The $\kvec$-derivative of the Bloch wavefunctions can be evaluated as
a covariant derivative,~\cite{sai02} or by \mbox{$\kvec\cdot\pvec$}
perturbation theory.~\cite{kdotp}

Equations~(\ref{eq:Hae_conv}) and (\ref{eq:M_periodic}) are adequate to
evaluate the NMR shielding tensor Eq.~(\ref{eq:converse}) in the context
of an all-electron method (such as FLAPW, or local-basis methods), in
which the interaction between core and valence electrons is treated explicitly.
However, in a pseudopotential framework, where the effect of the
core electrons has been replaced by a smooth effective potential,
Eqs.~(\ref{eq:Hae_conv}) and (\ref{eq:M_periodic}) are not sufficient
to evaluate the NMR shielding tensor.  One reason for this is that the
valence wave functions have been replaced by smoother pseudo wave functions which
deviate significantly from the all-electron ones in the core region.

In the following sections, we derive the formulas needed to calculate
the converse NMR shielding tensor in Eq.~(\ref{eq:converse}), in the
context of the pseudopotential method. Our derivation is based
on the GIPAW transformation~\cite{Pickard_Mauri_01} (see also
Appendix~\ref{app:gipaw}), that allows one to reconstruct all-electron
wave functions from smooth pseudopotential wave functions. For the sake of
simplicity, we assume all GIPAW projectors to be norm-conserving.

\subsection{The converse-NMR GIPAW hamiltonian}
In this section we derive the pseudopotential GIPAW hamiltonian
corresponding to the all-electron (AE) hamiltonian in Eq.~(\ref{eq:Hae_conv}). For
reasons that will be clear in the next section, we include an external
uniform magnetic field $\Bvec$ in addition to the magnetic field generated
by the point dipole $\mvec_s$. For the sake of simplicity, we carry out
the derivation for an isolated system (i.e.\ a molecule) in the symmetric
gauge $\Avec(\rvec)=(1/2)\Bvec\times\rvec$.  The generalization to
periodic systems is then performed at the end.

We start with the all-electron hamiltonian:
\beq\label{eq:Hae_conv2}
  \Hcal_\AE = \frac{1}{2} \left[\pvec + \frac{1}{c}
   \big(\Avec(\rvec)+\Avec_s(\rvec)\big) \right]^2 + V(\rvec)
\eeq
We now decompose Eq.~(\ref{eq:Hae_conv2}) in powers of $\Avec_s$ as
$\Hcal_\AE = \Hcal_\AE^{(s0)} + \Hcal_\AE^{(s1)} + \Hcal_\AE^{(s2)}$, where
\bea
  \Hcal_\AE^{(s0)} &=& \frac{1}{2}\Big(\pvec + \frac{1}{c}\Avec(\rvec)\Big)^2
  + V(\rvec) \label{eq:Hae_decomp_0} \\
  \Hcal_\AE^{(s1)} &=& \frac{1}{2c}\big( \pvec\cdot\Avec_s(\rvec) +
  \Avec_s(\rvec)\cdot\pvec + \frac{2}{c}\Avec(r)\cdot\Avec_s(\rvec) \big)
  \label{eq:Hae_decomp_1} \\
  \Hcal_\AE^{(s2)} &=& \frac{1}{2c^2}\Avec_s(\rvec)^2 \nonumber
\eea
We can neglect $\Hcal_\AE^{(s2)}$ in all calculations,
since $\mvec_s$ is a small perturbation to the electronic structure.

We then apply the GIPAW transformation Eq.~(\ref{eq:gipaw}) to the two
remaining terms, Eqs.~(\ref{eq:Hae_decomp_0}) and (\ref{eq:Hae_decomp_1}),
and we expand the results up to first order in the magnetic field.
\footnote{Note that the GIPAW transformation does not change despite the
fact that the vector potential for the converse method has changed and now
includes $\Avec_s$. The reason is that $\Avec_s$ is perpendicular to
the integration path of Eq.~(10) in Ref.~[\onlinecite{Pickard_Mauri_03}].}
At zeroth order in the external magnetic field $\Bvec$, the GIPAW
transformation of $\Hcal_\AE^{(s0)}$ and $\Hcal_\AE^{(s1)}$ yields the GIPAW
hamiltonian:
\bea\label{eq:H0_ps}
  \Hcal_\GIPAW &\equiv& \Hcal_\GIPAW^{(s0,0)} + \Hcal_\GIPAW^{(s1,0)} \\
  \Hcal_\GIPAW^{(s0,0)} &=& \frac{1}{2}\pvec^2 + V_\mathrm{loc}(\rvec) +
  \sum_\Rvec\VRNL \\
  \Hcal_\GIPAW^{(s1,0)} &=& \frac{1}{2}\left[\pvec\cdot\Avec_s(\rvec) +
    \Avec_s(\rvec)\cdot\pvec\right] + \sum_{\Rvec} \KRNL
\eea
where $V_\mathrm{loc}(\rvec)$ is the local Kohn-Sham potential and $\VRNL$
is the non-local pseudopotential in the separarable Kleinmann-Bylander (KB)
form:
\beq
  \VRNL = \sum_{nm}\ket{\beta_{\Rvec,n}}v_{\Rvec,nm}\bra{\beta_{\Rvec,m}}
\eeq
Similar to $\VRNL$, the term $\KRNL$ has the form of a non-local operator
\bea\label{eq:paramag_first}
  \KRNL &=& \frac{1}{2c}\sum_{nm}\ket{\ptilde_{\Rvec,n}}k_{\Rvec,nm}
  \bra{\ptilde_{\Rvec,m}} \\
  k_{\Rvec,nm} &=& \braket{\phi_{\Rvec,n}|\pvec\cdot\Avec_s(\rvec) +
    \Avec_s(\rvec)\cdot\pvec|\phi_{\Rvec,m}} -\nonumber\\
  && \braket{\phitilde_{\Rvec,n}|\pvec\cdot\Avec_s(\rvec) +
    \Avec_s(\rvec)\cdot\pvec|\phitilde_{\Rvec,m}}
\eea
The index $\Rvec$ runs over all atoms in the system, and the indexes $n$
and $m$, individually run over all projectors associated with atom $\Rvec$.
For a definition of $\ket{\phi_{\Rvec,n}}$ and $\ket{\phitilde_{\Rvec,n}}$
see Appendix~\ref{app:gipaw}.
Note that the set of GIPAW projectors $\ket{\ptilde_\Rvec}$ need not
be the same as the KB projectors $\ket{\beta_\Rvec}$. For
instance, in the case of norm-conserving pseudopotentials, one KB
projector per non-local channel is usually constructed. Conversely,
two GIPAW projectors for each angular momentum channel are usually
needed.

Equation~(\ref{eq:H0_ps}) is the Hamiltonian to be implemented in
order to apply a point magnetic dipole to the system. The first term of
$\Hcal_\GIPAW^{(s1,0)}$ can be applied to a wave function in real space
or in reciprocal space. The second term acts on the wave functions like
an extra non-local term and requires very little change to the
existing framework that applies the non-local potential.

At the first order in the magnetic field, the GIPAW transformation yields
two terms:
\bea\label{eq:H1_ps}
  \Hcal_\GIPAW^{(s0,1)} &=& \frac{1}{2c}\Bvec\cdot\Big(\Lvec + 
    \sum_\Rvec \Rvec\times\frac{1}{i} \left[\rvec,\VRNL\right]\Big) \\
  \Hcal_\GIPAW^{(s1,1)} &=& \frac{1}{2c}\Bvec\cdot\Big( \rvec\times\Avec_s(\rvec)
  +\sum_\Rvec \ERNL + \nn
  && {} + \sum_\Rvec \Rvec\times\frac{1}{i}
  \left[\rvec,\KRNL\right]\Big)
\eea
where $\Lvec = \rvec\times\pvec$ and $\ERNL$ is the non-local operator
\bea\label{eq:diamag_first}
  \ERNL &=& \sum_{nm}\ket{\ptilde_{\Rvec,n}}{\bf e}_{\Rvec,nm}\bra{\ptilde_{\Rvec,m}}\\
  {\bf e}_{\Rvec,nm} &=& \braket{\phi_{\Rvec,n}|(\rvec-\Rvec)\times\Avec_s(\rvec)|
  \phi_{\Rvec,m}} - \nonumber\\
  &&\braket{\phitilde_{\Rvec,n}|(\rvec-\Rvec)\times\Avec_s(\rvec)
  |\phitilde_{\Rvec,m}}
\eea
The two equations above will be used in the next section, in conjunction
with the Hellmann-Feynman theorem, to derive the GIPAW form of the
orbital magnetization.

\subsection{The orbital magnetization in the GIPAW formalism}
The orbital magnetization for a non spin-polarized system
is formally given by the Hellmann-Feynman theorem as
\beq
  \Mvec = -\Braket{\frac{\partial\Hcal_\AE}{\partial\Bvec}}_{B=0}
\eeq
In the GIPAW formalism this expectation value can be expressed
in terms of the GIPAW Hamiltonian and pseudo wave functions
\beq
  \Mvec = -\Braket{\frac{\partial\Hcal_\GIPAW}{\partial\Bvec}}_{B=0}
\eeq
By using the results of the previous section we find:
\begin{multline}\label{eq:M_from_HF}
  \Mvec = -\frac{1}{2c}\sum_n^\mathrm{occ}\Big\langle\psi_n\Big| 
  \Lvec + \rvec\times\Avec_s(\rvec) + \sum_\Rvec \ERNL +\\
  + i\sum_\Rvec \Rvec\times\left[\VRNL+\KRNL,\rvec\right]
  \Big|\psi_n\Big\rangle
\end{multline}
Note that in the expression above, $\psi_n$ are the eigenstates of
the GIPAW Hamiltonian in absence of any external magnetic fields.

While the formula~(\ref{eq:M_from_HF}) for $\Mvec$ can directly be
applied to atoms and molecules, it is ill-defined in the context of
periodic systems, owing to the presence of the position
operator---explicitly as in $\rvec$, but also implicitly as in $\Lvec$.
This problem can be remedied by applying the modern theory of orbital
magnetization.~\cite{Resta_05,Thonhauser_06,Ceresoli_06,Niu} The
goal is thus to reformulate Eq.~(\ref{eq:M_from_HF}) in terms of
$\braket{\rvec\times\vvec_\GIPAW}$. We can calculate this operator as
\begin{multline}\label{eq:rtimesvps}
  \rvec\times\vvec_\GIPAW = \rvec\times\frac{1}{i}
  [\rvec,\Hcal_\GIPAW]_{B=0} =\\
  = \Lvec + \rvec\times\Avec_s(\rvec) + 
  i\sum_\Rvec \rvec\times[\VRNL+\KRNL,\rvec]
\end{multline}
Replacing $\Lvec$ in Eq.~(\ref{eq:M_from_HF}) by the corresponding
expression calculated from Eq.~(\ref{eq:rtimesvps}), and regrouping the
terms, we obtain the central result of this paper:
\bea
  \Mvec &=& \Mvec_\mathrm{bare} + \Mvec_\mathrm{NL} + 
            \Mvec_\mathrm{para} + \Mvec_\mathrm{dia} \label{eq:M_GIPAW}\nn
  \Mvec_\mathrm{bare} &=& -\frac{1}{2c}\braket{\rvec\times\vvec_\GIPAW}
    \label{eq:M_bare}\\
  \Mvec_\mathrm{NL} &=& -\frac{i}{2c} \Braket{\sum_\Rvec (\Rvec-\rvec)\times
    \left[(\Rvec-\rvec),\VRNL\right]} \label{eq:M_NL}\\
  \Mvec_\mathrm{para} &=& -\frac{i}{2c} \Braket{\sum_\Rvec (\Rvec-\rvec)\times
     \left[(\Rvec-\rvec),\KRNL\right]} \label{eq:M_para}\\
  \Mvec_\mathrm{dia} &=& -\frac{1}{2c}\Braket{\sum_\Rvec \ERNL}
    \label{eq:M_dia}
\eea
where $\braket{\dots}$ stands for
$\sum_n^\mathrm{occ}\braket{\psi_n|\dots|\psi_n}$. The naming of
the various terms are in analogy to Ref.~[\onlinecite{Pickard_Mauri_01}].
The set of equations (\ref{eq:M_bare})--(\ref{eq:M_dia}) are now
valid both in isolated and periodic systems, as shown in detail in
Appendix~\ref{app:periodic}.

\section{Implementation and Computational Details}
\label{sec:implementation}
We have implemented the converse NMR method and its GIPAW transformation
into {\sc PWscf}, which is part of the {\sc Quantum-Espresso}
package.~\cite{pwscf}

In principle, the calculation of the NMR shielding is performed the
following way: The vector potential corresponding to the microscopic
dipole is included in the Hamiltonian and the Kohn-Sham equation is
solved self-consistently under that Hamiltonian. In practice, however,
in a first step one can equally as well find the ground state of the
unperturbed system.  Based on this ground state, in the second step
one can then introduce the dipole perturbation and reconverge to the
new ground state.  Note that the reconvergence of the small dipole
perturbation is usually very fast and only a small number of SCF steps
is necessary in addition to the ground state calculation. In fact, tests
have shown that a reconvergence is not even necessary---diagonalizing the
perturbed Hamiltonian only once with the unperturbed wave functions gives
results for the NMR shielding within 0.01 ppm of the fully converged
solution. This yields a huge calculational benefit for large systems,
where we calculate the unperturbed ground state once and then, based on
the converged ground-state wave functions, calculate all shieldings of
interest by non-SCF calculations.

In order to study the convergence of the NMR chemical shielding with
several parameters, we performed simple tests on a water molecule in
the gas phase. The molecule was relaxed in a box of 30 Bohr;
for all our calculations we used Troullier-Martin norm-conserving pseudopotentials~\cite{TM}
and a PBE exchange-correlation functional.~\cite{PBE}

First, we tested the convergence of the NMR shielding with respect to
the kinetic-energy cutoff $E_\text{kin}$ and the results are presented
in Table~\ref{tab:E_kin}. The shielding $\sigma$ is converged to within
0.02 ppm for a kinetic-energy cutoff of 80 Ryd. Similar tests on other
structures show similar results.

\begin{table}
\caption{\label{tab:E_kin}Chemical shielding $\sigma$ in ppm of a hydrogen atom in a water
molecule as a function of the kinetic-energy cutoff $E_{\text{kin}}$
(in units of Rydberg).}
\begin{tabular*}{\columnwidth}{@{}c@{\extracolsep{\fill}}ccc@{}}
\hline\hline
$E_\text{kin}$ & $\sigma$ & $E_\text{kin}$ & $\sigma$\\
(Ry)           & (ppm)    & (Ry)           & (ppm) \\
\hline
30 & 31.0009 &  70 & 31.1177\\
40 & 31.0595 &  80 & 31.1301\\
50 & 31.0637 &  90 & 31.1228\\
60 & 31.0832 & 100 & 31.1119\\
\hline\hline
\end{tabular*}
\end{table}

Next, we tested the convergence of the NMR shielding with respect to
the magnitude of the microscopic dipole $|\mvec_s|$ used and the energy
convergence criterion. At first sight it might appear difficult to
accurately converge the electronic structure in the presence of a
small microscopic magnetic point dipole. Thus, we tested using
different magnitudes for the microscopic dipole spanning several orders
of magnitude from $10^{-5}$ $\mu_B$ (which is actually much less than
the value of a core spin) to $10^3$ $\mu_B$ (which is obviously much
more than an electron spin). On the other hand, the ability to converge
the electronic structure accurately goes hand in hand with the energy
convergence criterion $E_{\text{conv}}$ that is used in such calculations.
This criterion is defined such that the calculation is considered
converged if the energy difference between two consecutive SCF steps is
smaller than $E_{\text{conv}}$. The results for the shielding as a
function of $|\mvec_s|$ and $E_{\text{conv}}$ are collected in
Table~\ref{tab:E_conv}. It is interesting to see that it is just as
simple to converge with a small dipole than it is to converge with
a large dipole. In either case, using at least $E_{\text{conv}}=10^{-4}$
Ryd yields results converged to within 0.1 ppm. Such a convergence criterion
is not even particularly ``tight'' and most standard codes use at least
$E_{\text{conv}}=10^{-6}$ Ryd as default. Note that first signs of
non-linear effects appear if large dipoles such as $|\mvec_s|=100\,\mu_B$
or $|\mvec_s|=1000\,\mu_B$ are used.
In conclusion of the above tests, we use $|\mvec_s|=1\,\mu_B$,
$E_{\text{conv}}=10^{-8}$ Ryd, and $E_{\text{kin}}=100$ Ryd for
all calculations.
\begin{table*}
\caption{\label{tab:E_conv}Chemical shielding in ppm of a hydrogen atom in a water
molecule.  The shielding is given as a function of the magnitude of the
microscopic dipole $|\mvec_s|$ (in units of Bohr magneton $\mu_B$)
and the energy convergence criterion $E_{\text{conv}}$ (in units of
Rydberg). $E_{\text{conv}}$ is defined such that the calculation is
considered converged if the energy difference between two consecutive
SCF steps is smaller than $E_{\text{conv}}$.}
\begin{tabular*}{\textwidth}{@{}l@{\extracolsep{\fill}}ccccccccr@{}}
\hline\hline
&\multicolumn{9}{c}{$E_{\text{conv}}$}\\
&\multicolumn{9}{c}{(Ry)}\\
$|\mvec_s|$ ($\mu_B$) &$10^{-2}$&$10^{-3}$&$10^{-4}$&$10^{-5}$&$10^{-6}$&$10^{-7}$&$10^{-8}$&$10^{-9}$& $10^{-10}$\\\hline
0.00001& 31.5170 & 31.2541 & 31.2541 & 31.1354 & 31.1403 & 31.1338 & 31.1356 & 31.1356 & 31.1356\\
0.0001 & 31.5265 & 31.2664 & 31.2664 & 31.1392 & 31.1390 & 31.1322 & 31.1300 & 31.1300 & 31.1300\\
0.001  & 31.5253 & 31.2667 & 31.2667 & 31.1395 & 31.1397 & 31.1328 & 31.1304 & 31.1304 & 31.1304\\
0.01   & 31.5251 & 31.2665 & 31.2665 & 31.1395 & 31.1394 & 31.1325 & 31.1303 & 31.1303 & 31.1303\\
0.1    & 31.5250 & 31.2664 & 31.2664 & 31.1393 & 31.1393 & 31.1323 & 31.1301 & 31.1301 & 31.1301\\
1.0    & 31.5250 & 31.2664 & 31.2664 & 31.1393 & 31.1393 & 31.1323 & 31.1301 & 31.1301 & 31.1301\\
10.0   & 31.5250 & 31.2663 & 31.2663 & 31.1392 & 31.1392 & 31.1322 & 31.1301 & 31.1301 & 31.1301\\
100.0  & 31.5212 & 31.2586 & 31.2586 & 31.1350 & 31.1327 & 31.1243 & 31.1256 & 31.1256 & 31.1252\\
1000.0 & 31.0167 & 30.5904 & 30.7197 & 30.6618 & 30.6334 & 30.6408 & 30.6403 & 30.6404 & 30.6405\\
\hline\hline
\end{tabular*}
\end{table*}

\subsection{Generation of GIPAW pseudopotentials}
\label{sec:pseudopotentials}
We have generated special-purpose norm-conserving pseudopotentials
for our GIPAW calculations. In addition to standard norm-conserving
pseudopotentials (PS), the GIPAW pseudopotentials include (i) the full
set of AE core atomic functions and (ii) the AE ($\phi_n$) and the PS
($\phitilde_n$) valence atomic orbitals.  The core orbitals contribution
to the isotropic NMR shielding is:
\beq
  \sigma_\mathrm{core} = \frac{1}{2 c} \sum_{n\in\mathrm{core}} 2 (2l_n+1)
  \Braket{\phi_n|\frac{1}{r}|\phi_n}
\eeq
The AE and PS valence orbitals are used to compute the coefficients
$k_{\Rvec,nm}$ and ${\bf e}_{\Rvec,nm}$ at the beginning of the calculation.
The PS valence orbitals are also used to compute the GIPAW projectors
$\ket{\ptilde_\Rvec}$ from:
\begin{gather}
  \ket{\ptilde_{\Rvec,n}} = \sum_m (S^{-1})_{nm} \ket{\phitilde_{\Rvec,m}}\\
  S_{nm} = \braket{\phitilde_{\Rvec,n}|\phitilde_{\Rvec,m}}_{R_c}
\end{gather}
$S$ is the overlap between atomic PS wave function, integrated up to the
cutoff radius of the corresponding pseudopotential channel.

We construct at least two projectors per angular momentum channel
by combining each valence orbital with one excited state with the same
angular momentum. For example, for hydrogen we include the $2s$
orbital in the set of atomic wave functions. For all second row elements,
we add the $3s$ and $3p$ orbitals, and so on. If any excited state turns out
to be unbound (as in the case of oxygen and fluorine), we generate an
atomic wave function as a scattering state at an energy 0.5 Ry higher
than the corresponding valence state. This procedure ensures 
that the GIPAW projectors are linearly independent and that the matrix
$S$ is not singular.

We found that the accuracy of the calculated NMR chemical shifts depends
critically on the cutoff radii of the pseudopotentials. Whereas the
total energy and the molecular geometry converge more quickly with respect
to reducing
the pseudopotential radii, the NMR chemical shift converges more slowely. Therefore,
GIPAW pseudopotentials have to be generated with smaller radii compared 
to the pseudopotentials usually employed for total energy
calculations. Table~\ref{tab:pseudo} reports the atomic configuration
and the cutoff radii used to generate the pseudopotentials.

\begin{table}
\caption{Electronic configuration and cutoff radii for the norm-conserving
pseudopotentials used in the present work.}
\begin{tabular*}{\columnwidth}{@{}lc@{\extracolsep{\fill}}ccc@{}}
\hline\hline
Atom & configuration & $r_c(s)$ & $r_c(p)$ & $r_c(d)$ \\
\hline
H    & $1s^2$          & 0.50 & $-$ & $-$\\
B    & [He] $2s^22p^1$ & 1.40 & 1.40 & $-$\\
C    & [He] $2s^22p^2$ & 1.50 & 1.50 & $-$\\
N    & [He] $2s^22p^3$ & 1.45 & 1.45 & $-$\\
O    & [He] $2s^22p^4$ & 1.40 & 1.40 & $-$\\
F    & [He] $2s^22p^5$ & 1.30 & 1.30 & $-$\\
P    & [Ne] $3s^23p^{2.5}3d^0$     & 1.90 & 2.10 & 2.10 \\
Si   & [Ne] $3s^23p^{1.3}3d^{0.2}$ & 2.00 & 2.00 & 2.00 \\
Cl   & [Ne] $3s^{1.75}3p^{4.5}3d^{0.25}$  & 1.40 & 1.40 & 1.40 \\
Cu   & [Ar] $4s^14p^03d^{10}$      & 2.05 & 2.20 & 2.05 \\
\hline\hline
\end{tabular*}
\label{tab:pseudo}
\end{table}

\section{Results}
\label{sec:results}
In this section we present results for molecules and solids. We first
calculated the absolute shielding tensor of some small molecules by
two different approaches, the direct (linear response) and the converse
method, in order to check that the two yield the same results.
Then, we compared the chemical shifts of fluorine compounds, calculated
by the converse method and by all-electron large basis set
quantum-chemistry calculations.
Finally, we report the calculated $^{29}$Si chemical shifts of
three SiO$_2$ polymorphs and the Cu shift of a
metallorganic compound.

\subsection{Small molecules}
We calculated the chemical shift of hydrogen, carbon, fluorine, phosphorus,
and silicon atoms of various small molecules.  First,
the structures were relaxed using PWSCF in a
box of 30 Bohr and a force convergence threshold of $10^{-4}$~Ry/Bohr.
Using the relaxed positions, the chemical shifts were calculated
using both the direct and converse method, and the results are shown in
Table~\ref{tab:simple_molecules}. This benchmark calculation shows that
the direct and the converse methods agree to within
less than 1\%.

\begin{table}
\caption{\label{tab:simple_molecules}Results for the absoule NMR shielding $\sigma$ in ppm of small molecules for the direct and converse methods. The core contribution to
the shielding is also shown.}
\begin{tabular*}{\columnwidth}{@{}l@{\extracolsep{\fill}}ccr@{}}\hline\hline
molecule & direct & converse & core\\\hline
{\bf H shielding}  & & & \\
CH$_4$       & 30.743 & 30.670 & 0.0 \\
C$_6$H$_6$   & 22.439 & 22.403 & 0.0 \\
SiH$_4$      & 27.444 & 27.413 & 0.0 \\
TMS          & 30.117 & 30.125 & 0.0 \\
{\bf C shielding}  & & & \\
CH$_4$       & 185.435 & 186.027 & 200.333 \\
C$_6$H$_6$   & 36.887 & 37.205 & 200.333 \\
CH$_3$F      & 93.704 & 94.250 & 200.333 \\
TMS          & 175.774 & 176.094 & 200.333 \\
{\bf F shielding}  & & & \\
CH$_3$F      & 448.562 & 447.014 & 305.815 \\
PF$_3$       & 277.148 & 275.819 & 305.815 \\
SiF$_4$      & 378.857 & 376.227 & 305.815 \\
SiH$_3$F     & 423.456 & 422.253 & 305.815 \\
{\bf P shielding}  & & & \\
P$_2$        & -323.566 & -320.201 & 908.854 \\
PF$_3$       & 150.603 & 150.856 & 908.854 \\
{\bf Si shielding} & & & \\
SiF$_4$      & 431.438 & 432.495 & 837.913 \\
SiH$_3$F     & 337.648 & 337.677 & 837.913 \\
Si$_2$H$_4$  & 230.830 & 230.489 & 837.913 \\
TMS          & 320.958 & 320.636 & 837.913 \\\hline\hline
\end{tabular*}
\end{table}

\subsection{Fluorine compounds}
In structural biology $^{19}$F NMR spectroscopy plays an important role
in determining the structure of protein
membranes.~\cite{maisch09}  The advantage over $^{15}$N and $^{17}$O
labeling is twofold: the natural abundance of $^{19}$F is nearly 100\%,
and $^{19}$F has spin $1/2$, i.e.\ a vanishing nuclear quadrupole
moment.
Quadrupole interactions in high-spin nuclei (e.g. $^{17}$O) are
responsible for the broadening of the NMR spectrum. On the contrary,
$^{19}$F NMR yields very sharp and resolved lines. In addition, it has
been found that the substitution of $-$CH$_3$ groups with $-$CF$_3$
in some amino-acids does not perturb the structure and the activity
of protein membranes, allowing for \emph{in vivo} NMR measurements.

In order to benchmark the accuracy of our method, we calculated the
$^{19}$F chemical shifts of ten fluorine compounds utilizing the converse
method, and we compared our results to all-electron gaussian-basis set
calculations, as well as to experimental data.
The molecules were first relaxed with Gaussian03,~\cite{gaussian}
with the 6-311+g(2d,p) basis set at the B3LYP level. Then,
we calculated the IGAIM chemical shift with Gaussian03, with
the cc-pVTZ, cc-pVQZ, cc-pV5Z and cc-pV6Z basis sets.~\cite{emsl}

\begin{figure}\begin{center}
\includegraphics[width=\columnwidth]{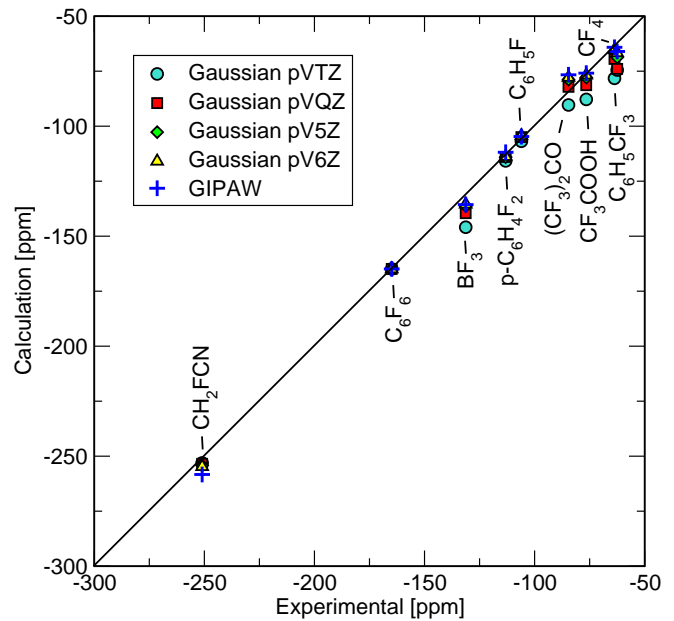}
\caption{(Color online) Experimental vs. calculated $^{19}$F chemical shifts in ppm,
with respect to CF$_3$Cl. For sake of clarity the chemical shift
of F$_2$ is not shown.}
\label{fig:fluorine}
\end{center}\end{figure}

To calculate the relative chemical shifts, we used C$_6$F$_6$ as a
secondary reference compound, and we used the experimental C$_6$F$_6$
chemical shift, to get the primary reference absolute shift (CF$_3$Cl). 
The results are shown in Table~\ref{tab:fluorine}
and in Fig.~\ref{fig:fluorine}. While compiling Table~\ref{tab:fluorine}
we suspected that the experimental values of p-C$_6$H$_4$F$_2$ and
C$_6$H$_5$F have been mistakenly exchanged in Ref.~[\onlinecite{19F_NMR}]. A
quick inspection of the original paper,~\cite{original_19F} confirmed our
suspicion.  The overall agreement of the converse method with experimental
data is very good, and of the same quality as the cc-pV6Z basis set,
which comprises 140 basis functions for 2nd row atoms. The calculation
time required by our plane-wave converse method is comparable to that of
cc-pV5Z calculations.

\begin{table*}
\caption{\label{tab:fluorine} Experimental and calculated $^{19}$F chemical shifts in ppm,
with respect to CF$_3$Cl. In all calculations (Gaussian03 and GIPAW plane
waves), we used C$_6$F$_6$ as the reference compound. The experimental
values of p-C$_6$H$_4$F$_2$ and C$_6$H$_5$F are exchanged with respect
to Ref.~[\onlinecite{19F_NMR}].  For molecules with inequivalent F atoms,
the average chemical shift is reported.  MAE is the mean absolute error
in ppm, with respect to experiment.}
\begin{tabular*}{\textwidth}{@{}l@{\extracolsep{\fill}}cccccc@{}}
\hline\hline
Molecule & Expt.~\cite{19F_NMR} & Gaussian cc-pVTZ & Gaussian cc-pVQZ & Gaussian cc-pV5Z & Gaussian cc-pV6Z & GIPAW converse\\
\hline
CH$_2$FCN         & --251    & --253.07 & --253.47 & --254.25 & --254.79 & --258.31 \\
C$_6$F$_6$        & --164.9  & --164.90 & --164.90 & --164.90 & --164.90 & --164.90 \\
BF$_3$            & --131.3  & --145.93 & --139.55 & --136.37 & --135.49 & --135.65 \\
p-C$_6$H$_4$F$_2$ & --113.15 & --115.77 & --113.98 & --114.07 & --113.78 & --111.84 \\
C$_6$H$_5$F       & --106    & --106.84 & --104.94 & --104.83 & --104.35 & --104.68 \\
(CF$_3$)$_2$CO    &  --84.6  &  --90.37 &  --82.12 &  --78.81 &  --77.83 &  --76.63 \\
CF$_3$COOH        &  --76.55 &  --87.82 &  --81.38 &  --77.88 &  --76.80 &  --75.90 \\
C$_6$H$_5$CF$_3$  &  --63.72 &  --78.24 &  --69.42 &  --66.21 &  --65.28 &  --64.16 \\
CF$_4$            &  --62.5  &  --74.48 &  --73.94 &  --68.76 &  --66.66 &  --66.05 \\
F$_2$             & 422.92  &  367.92 &  375.36 &  383.03 &  385.72 &  390.02 \\
\hline
MAE               &  $-$    &  13.19 &     9.40 &    7.35 &    6.69 &  5.64 \\
\hline\hline
\end{tabular*}
\end{table*}

\subsection{Solids}
In this section we present the $^{29}$Si and $^{17}$O chemical shifts
calculated by our converse method in four SiO$_2$ polymorphs: quartz,
$\beta$-cristobalite, coesite and stishovite. Coesite and stishovite are
metastable phases that form at high temperature and pressures
developed during a meteor impact.~\cite{NASA} Besides their
natural occurrence in meteors, they can also be artificially synthesized
by shock experiments.

We adopted the experimental crystal structures and atom positions
in all calculations. We used a cutoff of 100 Ryd and a k-point mesh of
8$\times$8$\times$8 for quartz, $\beta$-cristobalite and stishovite.
In the case of coesite, having the largest primitve cell (48 atoms), we
used a k-point mesh of 4$\times$4$\times$4.  In Table~\ref{tab:minerals}
we show a comparison between the calculated and the experimental
chemical shifts for the four crystals.  We determined the $^{29}$Si and
$^{17}$O reference shielding as the intercept of the least-square linear
interpolation of the $(\sigma_\mathrm{calc},\delta_\mathrm{expt})$ pairs.
Note that the nuclear magnetic dipole $\mvec_s$ breaks the symmetry
of the hamiltonian. Thus, we retained only the symmetry operations that
map site $s$ in $s'$ without changing the orientation of the magnetic
dipole (i.e. $s\rightarrow s', \mvec_s = \mvec_{s'}$). 

Another important point is that in periodic systems we are not just
including one nuclear dipole, but rather an infinite array. Thus, interactions
between $\mvec_s$ and its infinite periodic replicas become important,
and the chemical shift should be converged with respect to the supercell
size. To test for this convergence, we repeated the calculations for quartz and
$\beta$-cristobalite in a larger supercell and we found a change in chemical
shift of less than 0.1~ppm. This rapid convergence is due to the $1/r^3$
decay of the magnetic dipole interactions.

\begin{table}
\caption{\label{tab:minerals}Calculated and experimental $^{29}$Si and $^{17}$O NMR
chemical shifts of four SiO$_2$ crystals in ppm. Experimental data was taken from
Refs.~[\onlinecite{NASA}] and~[\onlinecite{profeta03}]. In the case of
coesite all inequivalent chemical shifts are reported.}
\begin{tabular*}{\columnwidth}{@{}l@{\extracolsep{\fill}}c@{\hspace{1cm}}c@{}}
\hline\hline
Mineral &  Calc. $\delta$ [ppm] & Expt. $\delta$ [ppm]\\
\hline
        & \multicolumn{2}{c}{\bf $^{29}$Si} \\
quartz               & $-$107.10 & $-$107.73 \\
$\beta$-cristobalite & $-$108.78 & $-$108.50 \\
stishovite           & $-$184.13 & $-$191.33 \\
coesite              & $-$107.30 & $-$107.73 \\
                     & $-$113.35 & $-$114.33 \\
        & \multicolumn{2}{c}{\bf $^{17}$O} \\
quartz               & 43.52     & 40.8 \\
$\beta$-cristobalite & 40.35     & 37.2 \\
stishovite           & 116.35    & N/A  \\
coesite              & 26.35     & 29   \\
                     & 39.66     & 41   \\
                     & 52.70     & 53   \\
                     & 56.84     & 57   \\
                     & 59.03     & 58   \\
\hline\hline
\end{tabular*}
\end{table}

\subsection{Large Systems}
\begin{figure}\begin{center}
\includegraphics[width=0.8\columnwidth,clip=true]{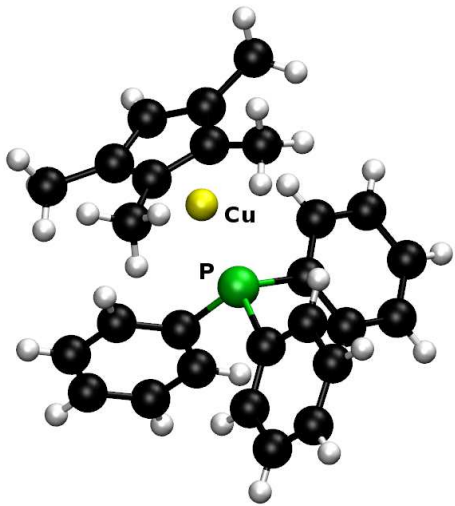}\vspace{5mm}
\caption{(Color online) The molecular structure of the metallocene,
tetramethylcyclopentadienyl copper(I) triphenylphosphine
(CpCuPPh3).  The copper is shown in gold (light gray) forming the 
metallocene bond, while the phosphorus is green (gray).  The 
crystal structure of this materials consists of a 228-atom
orthorhombic unit cell. }
\label{fig:cpcupph3}
\end{center}\end{figure}

Reactive sites in biological systems such as organometallic molecules,
as well as inorganic materials, are of great importance. In particular,
there is a surge of interest in studying copper(I) reactive sites
using solid-state NMR. NMR experiments on these materials are challenging 
because of the large nuclear quadrupole moments of $^{63}$Cu and $^{65}$Cu.
Here, we present the results for the copper-phosphine metallocene,
tetramethylcyclopentadienyl copper(I) triphenylphosphine (CpCuPPh3),
which as a solid contains 228 atoms in a primitive orthorhombic unit
cell.  The molecular structure is shown in Fig.~\ref{fig:cpcupph3}.
The properties of the shielding tensor for the copper environment were
observed experimentally for the solid material, and simulated using
quantum-chemical methods on the molecular complex.~\cite{tang07}

While the converse approach can calculate the chemical shift for this
large system easily, it is more challenging for the linear-response
method, which in our experience took much longer in general, did not
finish at all, or was unable to handle such large systems.
We calculated the copper chemical shift for CpCuPPh3 using the converse
method with an energy cutoff of 80 Ry in the self-consistent step and
PBE pseudopotentials.  While previous quantum-chemical calculations were
able to reasonably reproduce the experimental span (1300 ppm) and the skew
(0.95) of the chemical shielding tensor, they were not able to calculate
the chemical shift itself (0~ppm relative to copper (I) chloride),
with an inaccuracy of several hundred ppm.~\cite{tang07}  In addition
to yielding excellent agreement with experiment to within 2 ppm for the
chemical shift, our calculations also gave good results for the span 
(1038 ppm) and the skew (0.82) of the chemical shielding tensor.

\section{Discussion}\label{sec:discussion}
The results presented in the previous sections and in a previous
work~\cite{Thonhauser_09} show that DFT is able to predict accurately
the chemical shift of molecules and solids. In general, we expect
this to be true for any weakly-correlated system, well described by
the generalized-gradient approximation (GGA).  In addition to that,
relativistic corrections to the NMR chemical shifts are negligible for
all light elements in the periodic table, and become important starting
from 4th row elements.

However, there will always exist ``difficult'' cases in which relativistic
corrections cannot be neglected and/or one has to go beyond DFT with
standard local functionals. This is an active
field of research in quantum chemistry~\cite{NMR-dirac,NMR-beyond-DFT}
and today it is customary to compute NMR chemical shifts with semi-local
hybrid DFT functionals (such as B3LYP). Most quantum-chemistry codes
allow the inclusion of relativistic effects (spin orbit) by perturbation
theory; furthermore, fully-relativistic (four component) solutions of
the Dirac-Breit equation have recently been implemented.~\cite{NMR-dirac}

To the best of our knowledge, all existing ab-initio codes, calculate
NMR shifts by perturbation theory.
Among them, localized-basis sets
are the most popular choice to expand wave functions. This
leads to very complicated mathematical expressions and to 
gauge-dependent results. Only two plane-wave, linear-response
implementations~\cite{Pickard_Mauri_01,Sebastiani_01} have been
reported. Our converse-NMR method is built on Mauri's GIPAW method,
but has the advantage of circumventing the need for a linear response
framework.

The main advantage of our converse-NMR method is that it requires
only the ground state wave functions and Hamiltonian to calculate the
orbital magnetization. Since no external magnetic field is included
in the calculation, our method solves the gauge-origin problem.  Moreover,
``difficult cases'' can be treated easily by our converse method,
provided that relativistic corrections and many-body effects are
included in the Hamiltonian. Thus, one can concentrate effectively all
efforts in developing advanced post-DFT theories (i.e. DFT+U, DMFT,
hybrid functionals, self-interaction-free methods) and benchmark them
against NMR experiments.

\section{Summary}
\label{sec:summary}
In this paper we have generalized the recently developed converse NMR
approach~\cite{Thonhauser_09} such that it can be used in conjunction
with norm-conserving, non-local pseudopotentials.  We have tested our
approach both in finite and periodic systems, on small molecules, four
silicate minerals and a molecular crystal.  In all cases, we have found
very good agreement with established methods and experimental results.

The main advantage of the converse-NMR method is that it requires only
the ground state wave functions and Hamiltonian, circumventing the need
of any linear response treatment.  This is of paramount importance for
the rapid development and validation of new methods that go beyond DFT.

Currently, we are applying the converse NMR method to study large
biological systems such as nuclei acids\cite{Cooper_08}
and drug-DNA interactions\cite{Cooper2_08,Li_09} in conjunction with a recently
developed van der Waals exchange-correlation functional.\cite{Thonhauser_07, Langreth_09}
We are also exploring the
possibility to calculate non-perturbatively the Knight shift in metals.
Finally, the converse method can be used to calculate the EPR g-tensor
in molecules and solids.~\cite{ceresoli-EPR}

\section{Acknowledgments}
This work was supported by the DOE/SciDAC Institute on Quantum Simulation
of Materials and Nanostructures.  All computations were performed on the
Wake Forest University DEAC Cluster with support from the Wake Forest
University Science Research Fund.  D.\,C.\ acknowledges partial support
from ENI.

\appendix
\section{The GIPAW transformation}\label{app:gipaw}
The starting point is the projector augmented wave (PAW) transformation:~\cite{Blochl_94}
\beq
  \mathcal{T} = 1 + \sum_{\Rvec,n} \left( \ket{\phi_{\Rvec,n}} -
  \ket{\phitilde_{\Rvec,n}}\right) \bra{\ptilde_{\Rvec,n}}
\eeq
which connects an all-electron wave function $\ket{\psi}$ to the
corresponding pseudopotential wave function $\ket{\psitilde}$ via:
$\ket{\psi}=\mathcal{T}\ket{\psitilde}$.  Here, $\ket{\phi_{\Rvec,n}}$
are all-electron partial waves, $\ket{\phitilde_{\Rvec,n}}$ are
pseudopotential partial waves, and $\bra{\ptilde_{\Rvec,n}}$ are PAW
projectors. The sum runs over the atom positions $\Rvec$. $n$ is a combined
index that runs over the set of projectors attached to atom $\Rvec$.
In the original PAW formalism, there are two sets of projectors per
angular momentum channel ($0,\dots,l_\mathrm{max}$), each with $(2l+1)$
projectors for a total of $2(l_\mathrm{max}+1)^2$ PAW projectors.

The expectation value of an all-electron operator $O_\AE$ between
all-electron wave functions can then be expressed as the expectation
value of a pseudo operator $O_\PS$ between pseudo wave functions as
$\braket{\psi|O_\AE|\psi} = \braket{\psitilde|O_\PS|\psitilde}$, where
the $O_\PS$ is given by:
\begin{multline}\label{eq:paw}
  O_\PS \equiv \mathcal{T}^+ O_\AE \mathcal{T} = \\
  = O_\AE + \sum_{\Rvec,nm} \ket{\ptilde_{\Rvec,n}} \Big[
    \braket{\phi_{\Rvec,n}|O_\AE|\phi_{\Rvec,m}} - \\
  -\braket{\phitilde_{\Rvec,n}|O_\AE|\phitilde_{\Rvec,m}}
   \Big] \bra{\ptilde_{\Rvec,m}}
\end{multline}
In the presence of external magnetic fields the PAW transformation is no
longer invariant with respect to translations (except in the very simple
case of only one augmentation region). This deficiency was resolved by
Mauri \emph{et al.} who developed the GIPAW (``gauge including PAW'')
method~\cite{Pickard_Mauri_01}, which is similar to the PAW
transformation from Eq.~(\ref{eq:paw}) but with the inclusion of phase
factor compensating the gauge term arising from the translation of
a wave function in a magnetic field. The GIPAW transformation in the symmetric
gauge reads:
\begin{multline}
  \mathcal{T}_G = 1 + \sum_{\Rvec,n} e^{\exponent} \left( 
  \ket{\phi_{\Rvec,n}} - \ket{\phitilde_{\Rvec,n}}\right)\cdot\\
  \bra{\ptilde_{\Rvec,n}} e^{-\exponent}
\end{multline}
and the corresponding pseudopotential operator $O_\PS$ is:
\begin{multline}\label{eq:gipaw}
  O_\PS \equiv {\mathcal{T}_G}^+ O_\AE \mathcal{T}_G = O_\AE +
    \sum_{\Rvec,nm}e^{\exponent}\ket{\ptilde_{\Rvec,n}} \cdot\\
  \Big[\braket{\phi_{\Rvec,n}|e^{-\exponent}\,O_\AE\,e^{\exponent}|\phi_{\Rvec,m}} - \\ 
  -\braket{\phitilde_{\Rvec,n}|e^{-\exponent}\,O_\AE\,e^{\exponent}|\phitilde_{\Rvec,m}}\Big] \\
  \cdot\bra{\ptilde_{\Rvec,m}}e^{-\exponent}
\end{multline}
In the following we will refer to the often occurring part
$\hat{O}_\AE=e^{-\exponent}\,O_\AE\,e^{\exponent}$ as \emph{inner
operator} and denote it with a hat. The above expression allows
us to calculate accurate expectation values of operators within a
pseudopotential approach. In this work we carry out the derivation
working in the symmetric gauge $\Avec(\rvec)=(1/2)\Bvec\times\rvec$.
This is not an issue, since all physical quantities we are working with,
are gauge invariant.

One useful property of the GIPAW transformation is:
\begin{multline}\label{eq:identity}
  e^{-\exponent}\Big[\pvec + \frac{1}{c}\Avec(\rvec)\Big]^n e^{\exponent} =\\
  = \Big[\pvec + \frac{1}{c}\Avec(\rvec-\Rvec)\Big]^n
\end{multline}

\section{Periodic systems}\label{app:periodic}
Note that the set of equations~(\ref{eq:M_GIPAW}) are well
defined also in periodic boundary conditions. In fact,
$\Mvec_\mathrm{bare}$ can be calculated by the Modern Theory of the
Orbital Magnetization~\cite{Resta_05,Thonhauser_06,Ceresoli_06,Niu} as:
\begin{multline}
  \Mvec_\mathrm{bare} = -\frac{1}{2c}\mathrm{Im} \sum_{n\kvec}
  f(\epsilon_{n\kvec})\bra{ \partial_\kvec u_{n\kvec}}\\
  \times\left(\Hcal_\kvec + \epsilon_{n\kvec} -
  2\epsilon_\mathrm{F}\right)\ket{\partial_\kvec u_{n\kvec}}\Big|_\mathrm{GIPAW}
\end{multline}
where $\Hcal_\kvec$ is the GIPAW hamiltonian, $\epsilon_{n\kvec}$ and
$\ket{u_{n\kvec}}$ are its eigenvalues and eigenvectors.

The position operator appearing in the other terms in
Eq.~(\ref{eq:M_GIPAW}) is well defined because the projectors
$\ket{\beta_{\Rvec,n}}$ and $\ket{\ptilde_{\Rvec,n}}$ are non-vanishing
only inside an augmentation sphere, centered around atom $\Rvec$.

The expression of $\Mvec_\mathrm{NL}, \Mvec_\mathrm{para}$ and
$\Mvec_\mathrm{para}$ can be further manipulated in order to work
with Bloch wave functions. In the following part, we show it only for
$\Mvec_\mathrm{NL}$, since $\Mvec_\mathrm{para}$ is similar. The term
$\Mvec_\mathrm{dia}$ can be manipulated instead in a trivial way.

\begin{widetext}
Let's consider the expectation value of
\beq
  (\rvec-\Rvec)\times\left[\rvec-\Rvec,\VRNL\right] =
    -(\rvec-\Rvec)\times\left(\VRNL\right)\,(\rvec-\Rvec)
\eeq
on a Bloch state $\ket{\psi_{n\kvec}}=e^{i\kvec\rvec}\ket{u_{n\kvec}}$:
\beq
  -\sum_\Rvec \braket{u_{n\kvec}|e^{-i\kvec\rvec} (\rvec-\Rvec)
    \times(\VRNL)\,(\rvec-\Rvec)e^{i\kvec\rvec}|u_{n\kvec}} =
\eeq
\beq
  =-\sum_{\Lvec\tauvec}\sum_{ij}
  \braket{u_{n\kvec}|e^{-i\kvec\rvec} (\rvec-\Lvec-\tauvec)
    \ket{\beta_{\Lvec+\tauvec,i}}
    \times v^{L+\tau}_{ij} \bra{p_{\Lvec+\tauvec,j}}
    (\rvec-\Lvec-\tauvec)e^{i\kvec\rvec}|u_{n\kvec}}\;,
\eeq
where $\Lvec$ are the real space lattice vectors (not to be confused
with the angular momentum operator) and $\tauvec$ are the position of
the atoms in the unit cell. Inserting two canceling phase factors:
\beq
  \cdots=-\sum_{\Lvec\tau}\sum_{ij}
  \braket{u_{n\kvec}|e^{-i\kvec(\rvec-\Lvec-\tauvec)} (\rvec-\Lvec-\tauvec)
    \ket{\beta_{\Lvec+\tauvec,i}}
    \times v^{L+\tau}_{ij} \bra{\beta_{\Lvec+\tauvec,j}}
    (\rvec-\Lvec-\tauvec)e^{i\kvec(\rvec-\Lvec-\tauvec)}|u_{n\kvec}}
\eeq
one can recognize immediately the $\kvec$ derivative of the KB projectors.
In addition, since the KB projectors vanish outside their
augmentation regions, it is possible to insert a second sum over $\Lvec'$
running on the right hand side of the cross product:
\beq
  \cdots=-\sum_{\Lvec\Lvec'\tau}\sum_{ij}
  \braket{u_{n\kvec}|(-1/i)\partial_\kvec\left(e^{-i\kvec(\rvec-\Lvec-\tauvec)}
  \ket{\beta_{\Lvec+\tauvec,i}}\right)
  \times v^{L+\tau}_{ij}\, (1/i)\partial_\kvec
  \left(\bra{\beta_{\Lvec'+\tauvec,j}}e^{i\kvec(\rvec-\Lvec-\tauvec)}\right)
  |u_{n\kvec}}
\eeq
\beq
  =-\sum_{\tau}\sum_{ij}\braket{u_{n\kvec}|
  \partial_\kvec\left(\sum_\Lvec e^{-i\kvec(\rvec-\Lvec-\tauvec)}
    \ket{\beta_{\Lvec+\tauvec,i}}\right)
    \times v^{L+\tau}_{ij}\,
  \partial_\kvec\left(\sum_{\Lvec'}\bra{\beta_{\Lvec'+\tauvec,j}}
    e^{i\kvec(\rvec-\Lvec'-\tauvec)}\right)
  |u_{n\kvec}}
\eeq
In periodic systems the structure factors can be absorbed by the projectors:
\bea
  V_\mathrm{NL}^\kvec &=& \sum_{\tauvec}\sum_{ij}
  \ket{\beta_{\tauvec,i}^\kvec} v_{\tau,ij} \bra{\beta_{\tauvec,j}^\kvec}\\
  \ket{\beta_{\tauvec,i}^\kvec} &=& \sum_\Lvec e^{-i\kvec(\rvec-\Lvec-\tauvec)}
  \ket{\beta_{\Lvec+\tauvec,i}}
\eea
Finally:
\begin{subequations}
\bea
  \Mvec_\mathrm{NL} &=& \frac{1}{2c} \sum_{n\kvec}^\mathrm{occ}
    \sum_{\tauvec,ij} \frac{1}{i}
    \Braket{u_{n\kvec}|\partial_\kvec\beta_{\tauvec,i}^\kvec}
    \times v_{\tau,ij}\,
    \Braket{\partial_\kvec\beta_{\tau,j}^\kvec|u_{n\kvec}} \\
  \Mvec_\mathrm{para} &=& \frac{1}{2c} \sum_{n\kvec}^\mathrm{occ}
    \sum_{\tauvec,ij} \frac{1}{i}
    \Braket{u_{n\kvec}|\partial_\kvec\ptilde_{\tauvec,i}^\kvec}
    \times k_{\tau,ij}\,
    \Braket{\partial_\kvec\ptilde_{\tau,j}^\kvec|u_{n\kvec}} \\
  \Mvec_\mathrm{dia} &=& -\frac{1}{2c}\sum_{n\kvec}^\mathrm{occ}\Braket{
     u_{n\kvec}|\ptilde_{\tauvec,i}^\kvec}
    \bm{e}_{\tau,ij}\,
    \Braket{\ptilde_{\tau,j}^\kvec|u_{n\kvec}}
\eea
\end{subequations}
\end{widetext}
This completes the main result and it allows us to calculate the orbital
magnetization in the presence of non-local pseudopotentials. With this
result we can now easily and efficiently calculate the NMR chemical
shift for elements heavier then hydrogen using Eq.~(\ref{eq:converse}).


\end{document}